\documentstyle[aps,prl,multicol,psfig]{revtex}
\begin{document}
\title{\bf Extremal Paths on a Random Cayley Tree}
\author{Satya N. Majumdar$^{1,2}$ and P. L. Krapivsky$^{1,3}$}
\address{$^1$Laboratoire de
   Physique Quantique (UMR C5626 du CNRS), Universit\'e Paul Sabatier,
   31062 Toulouse Cedex, France}
\address{$^2$Tata Institute of
   Fundamental Research, Homi Bhabha Road, Mumbai-400005, India}
\address{$^3$Center for Polymer Studies and Department of Physics, 
   Boston University, Boston, MA 02215, USA}

\maketitle

\begin{abstract}

\noindent
We investigate the statistics of extremal path(s) (both the shortest
and the longest) from the root to the bottom of a Cayley tree.  The
lengths of the edges are assumed to be independent identically
distributed random variables drawn from a distribution $\rho(l)$.
Besides, the number of branches from any node is also random.  Exact
results are derived for arbitrary distribution $\rho(l)$.  In
particular, for the binary $\{0,1\}$ distribution
$\rho(l)=p\delta_{l,1}+(1-p)\delta_{l,0}$, we show that as $p$
increases, the minimal length undergoes an unbinding transition from a
`localized' phase to a `moving' phase at the critical value,
$p=p_c=1-b^{-1}$, where $b$ is the average branch number of the tree.
As the height $n$ of the tree increases, the minimal length saturates
to a finite constant in the localized phase ($p<p_c$), but increases
linearly as $v_{\rm min}(p)n$ in the moving phase ($p>p_c$) where the
velocity $v_{\rm min}(p)$ is determined via a front selection
mechanism. At $p=p_c$, the minimal length grows with $n$ in an
extremely slow double logarithmic fashion. The length of the maximal
path, on the other hand, increases linearly as $v_{\rm max}(p)n$ for
all $p$. The maximal and minimal velocities satisfy a general duality
relation, $v_{\rm min}(p)+v_{\rm max}(1-p)=1$, which is also valid for
directed paths on finite-dimensional lattices.

\medskip\noindent
{PACS numbers: 05.40.-a, 64.60.Cn, 02.50.-r}
\end{abstract}

\begin{multicols}{2}
\section{Introduction}
Optimization lies at the heart of a vast number of
phenomena: Particles ``seek'' paths with minimal action, species
``try'' to maximize fitness, companies minimize cost.  Many
optimization problems, e.g., the traveling salesman problem, are
notoriously hard\cite{G+J}.  Additionally, optimization problems often
involve randomness which makes them even more complicated. When the
number of random entries is very large, however, optimization problems
might become simpler thanks to the ``concentration of the
measure''. This means that the probability distribution of some random
variable becomes highly localized, almost like a delta function
concentrated around its average value.  This phenomenon is well-known
in probability theory where, for instance, it accounts for the fact
that after flipping a coin $N$ times, the probability that the number
of heads exceeds $N/2$ by more than, say, $100\sqrt{N}$ is about
$10^{-800}$.  Below, we shall investigate an optimization problem in
random media which demonstrates this concentration of the measure in
the strongest sense (the variance of the optimized quantity is finite)
and displays an unbinding phase transition.  A remarkable hidden
connection with traveling wave phenomena allows to explain both the
concentration of the measure and the phase transition. 

The optimization problem considered in this work can be formulated as
follows: Take a rooted tree whose bonds have random lengths and find
descending paths of extremal total length.  We assume that a random
length (or energy) is assigned to each bond of the tree, and that
lengths are independent and chosen from the same probability
distribution $\rho(l)$.  The total length of a path is the sum of
lengths of the bonds along the path, and we want to determine the
minimal (maximal) length among the paths from the root to the bottom
of the tree.  We focus on a rooted Cayley tree where each node, except
the root, has coordination number 3; the coordination number of the
root is 2.  However, our main results can be easily generalized to the
case of a tree with arbitrary coordination number, and even to trees
where the coordination number at any node is random.

Similar problems arise in numerous fields ranging from computer
science\cite{karp} to condensed matter physics\cite{HHF,zhang}, where
it is known as the problem of directed polymers in random media.
Indeed, the minimal path problem can be considered as the zero
temperature limit of the problem of directed polymer on Cayley
trees. Directed polymers on Cayley trees have been investigated in the
past\cite{D+S,BPP}, and recently this problem has resurfaced in a
surprisingly large number of apparently unrelated
problems\cite{CdC,tang,C+D}.  However, the emphasis of
Refs.\cite{D+S,BPP,CdC,tang,C+D} was on the spin glass like transition
occurring at a finite temperature. Here, we consider exactly zero
temperature and focus on the unbinding transition driven by the
parameter $p$ of the bimodal distribution,
$\rho(l)=p\delta_{l,1}+(1-p)\delta_{l,0}$.  Recently, similar
unbinding (or ``depinning'') phase transitions have been found in a
number of non-equilibrium processes without quenched
disorder\cite{Hin,HM,GM,MKB}.

Let us first outline our main results. We derive exactly the
statistics of both the minimal and maximal path on a random Cayley
tree for arbitrary distribution $\rho(l)$ of the edge lengths. We find
a class of distributions for which the minimal path undergoes an
unbinding phase transition from a localized phase to a moving phase as
a parameter of the distribution is varied. A particular example, which
we study in detail in this paper, is the bimodal distribution.  In the
bimodal case, length of each bond can be either $1$ with probability
$p$ or $0$ with probability $(1-p)$.  For any path starting at the
root and moving down, the length of the path is the sum of the lengths
of the bonds along the path, i.e., the number of $1$'s along the path.
There are $2^n$ paths from the root to the $n^{\rm th}$ level. We show
that the minimal length exhibits an unbinding transition as the
parameter $p$ varies between 0 and 1.  There exists a critical value
$p_c=1/2$ such that the average minimal length grows linearly with $n$
when $p>1/2$, but gets pinned, i.e., remains finite for large $n$ when
$p<1/2$. More specifically, we find that for large $n$, the average
minimal length $\langle L_n^{\rm min}\rangle$ behaves as
\begin{equation}
\label{Ln}
\langle L_n^{\rm min}\rangle\simeq\cases
                    {v_{\rm min}(p)n                 &$p>1/2$,\cr
                              (\ln 2)^{-1}\ln\ln n   &$p=1/2$,\cr
                              {\rm finite}           &$p<1/2$.\cr}
\end{equation}
For $p>1/2$, we shall later compute the minimal ``velocity'' $v_{\rm
min}(p)$ exactly. The average maximal length, on the other hand,
behaves as
\begin{equation}
\langle L_n^{\rm max}\rangle\simeq 
v_{\rm max}(p)n  \qquad        $ {\rm for all }p$,
\label{lmax}
\end{equation}
where $v_{\rm max}(p)+v_{\rm min}(1-p)=1$. This duality relation is
very general and valid for directed paths on arbitrary graphs, i.e.,
for finite-dimensional lattices. We also derive a generalized duality
relation between minimal and maximal velocities for a certain class of
bounded distributions.

The minimal length problem on a rooted Cayley tree, where the
coordination number at each node is random, also displays an unbinding
transition at $p_c=1-b^{-1}$, where $b$ is the average number of
branches per node.  At the critical point, the average minimal length
$\langle L_n^{\rm min}\rangle$ exhibits the same double logarithmic
growth for any $b>1$, and above $p_c$ the minimal length grows
linearly with $n$ with velocity $v_{\rm min}(p,b)$.

The rest of this paper is organized as follows. In Section II, we
study the unbinding transition of the minimal path for the special
bimodal distribution in detail. In Section III, we study the maximal
path for the bimodal case and derive the duality relation. In Section
IV, we generalize the results to a Cayley tree with random branching.
In Section IV, we go beyond the bimodal distribution and obtain
generalized results for arbitrary distribution of the edge length.
Finally we conclude in Section VI with a brief summary and outlook.

\section{Minimal Path(s)}

Instead of studying $P_n^{\rm min}(x)={\rm Prob}(L_n^{\rm min}=x)$, it
proves convenient to consider the cumulative distribution,
$P_n(x)={\rm Prob}(L_n^{\rm min}\geq x)$. Clearly, $P_n(x)$ is the
probability that all possible paths to the $n^{\rm th}$ level have
lengths $\geq x$. Once we get $P_n(x)$, the distribution of the
minimal length is found from the obvious identity $P_n^{\rm
min}(x)=P_n(x)-P_n(x+1)$.

It is easy to see that $P_n(x)$ satisfies the following recursion relation
\begin{equation}
\label{Pn}
P_{n+1}(x)=\left[(1-p)P_n(x)+pP_n(x-1)\right]^2.
\end{equation}
This relation can be derived by analyzing various possibilities for
the lengths of the two edges issuing from the root and taking into
account that two subsequent daughter trees are statistically
independent.  For instance, both lengths are equal to zero with
probability $(1-p)^2$, and paths in the subsequent trees have lengths
$\geq x$ with probability $P_n^2(x)$; this provides the contribution
$(1-p)^2P_n^2(x)$ to $P_{n+1}(x)$.  Similarly, one gets the
contributions $2p(1-p)P_n(x-1)P_n(x)$ and $p^2P_n^2(x-1)$, which sum
up to Eq.~(\ref{Pn}).

We must solve  Eq.~(\ref{Pn}) subject to the initial condition 
\begin{equation}
\label{P0}
P_0(x)=\cases{1      &$x\leq 0$,\cr
              0      &$x>0$.\cr}
\end{equation}
Clearly, the minimal length $L_n^{\rm min}$ is a random variable which
takes values between $0$ and $n$, and Eqs.~(\ref{Pn})--(\ref{P0})
indeed show that $P_n(x)=1$ for $x\leq 0$ and $P_n(x)=0$ for $x>n$.

When $n$ grows, $L_n^{\rm min}$ increases as well and thus it should
reach a limit. For $p<1/2$, this limit is finite, i.e., one can find
an infinite path with only finite number of edges of length
1. Mathematically, it means that $P_n(x)$ approaches a stationary
distribution, $P_n(x)\to P(x)$, which satisfies
\begin{equation}
\label{P}
P(x)=\left[(1-p)P(x)+pP(x-1)\right]^2.
\end{equation}
Starting from $P(0)=1$, one computes $P(x)$ recursively:
$P(1)=p^2(1-p)^{-2}$, etc. $P(x)$ vanishes extremely fast in the large
$x$ limit: $\ln P(x)\sim 2^x \ln p$.

For $p\geq 1/2$, the minimal length grows as $n$ increases.  To
understand why this is so, recall that $P_n(0)\equiv 1$ and look at
$P_n(1)$. By inserting $P_n(1)=1-q_n$ into Eq.~(\ref{Pn}) we find
$q_{n+1}=2(1-p)q_n-(1-p)^2q_n^2$. Hence $P_n(1)\to 1$ for $p\geq 1/2$.
Proceeding this line of reasoning one can verify that $P_n(x)\to 1$
when $n\to\infty$.  Additionally, we see the difference between the
cases of $p>1/2$ and $p=1/2$: In the former situation, $P_n(x)\to 1$
exponentially fast, while in the latter situation the approach is
algebraic.  Therefore, the size of the region where $P_n(x)$ is close
to 1 should grow linearly with $n$ when $p>1/2$.  Thus the
distribution $P_n(x)$ approaches the traveling wave form (see Fig. (1)),
\begin{equation}
\label{trav}
P_n(x)\to {\cal P}^2(y), \quad y=x-\langle L_n^{\rm min}\rangle.
\end{equation}
In Eq.~(\ref{trav}), we have chosen $y=x-\langle L_n^{\rm min}\rangle$ as the
wave variable. One can justify this choice by noting that the relation
$P_n^{\rm min}(x)=P_n(x)-P_n(x+1)$ leads to identity
\begin{equation}
\label{ident}
\langle L_n^{\rm min}\rangle=\sum_{x=0}^\infty x\left[P_n(x)-P_n(x+1)\right]
                  =\sum_{x=1}^\infty P_n(x),
\end{equation}
and hence $\langle L_n^{\rm min}\rangle$ is indeed an appropriate
characterization of the location of the front.  By inserting the wave
form (\ref{trav}) into Eq.~(\ref{Pn}) and using $\langle L_n^{\rm
min}\rangle\simeq vn$ we find,
\begin{equation}
\label{Pwave}
{\cal P}(y-v)=(1-p){\cal P}^2(y)+p{\cal P}^2(y-1),
\end{equation}
The solution of this equation looks like a $[1-0]$ waveform with
$P(y)\to 1$ as $y\to -\infty$ and $P(y)\to 0$ as $y\to \infty$.  While
it is very hard to solve this nonlinear, nonlocal equation exactly,
the velocity can be found by analyzing the tail region, $y\to
-\infty$, where $1-{\cal P}(y)$ is small.  Linearizing
Eq.~(\ref{Pwave}) in this region and noting that it admits an
exponential solution, $1-{\cal P}(y)\sim e^{\lambda y}$, one finds
that the velocity $v_{\lambda}$ is related to the decay exponent
$\lambda$ via
\begin{equation}
\label{lambda}
v_{\lambda}=-{\ln\left[2(1-p)+2p\,e^{-\lambda}\right]\over \lambda}.
\end{equation}

When $\lambda>\ln{2p\over 2p-1}$, the velocity $v_{\lambda}$ is
positive.  While any such $\lambda$ is in principle allowed, and thus
the entire velocity range of is feasible, a particular value is
actually selected. This is similar to the velocity selection in a
large class of problems with a traveling wave
front\cite{front1,front2}. It is well known that for a wide class of
initial conditions, the extremum value is chosen. From this general
front selection principle one can infer that in the present case the
selected value $\lambda=\lambda^*(p)$ corresponds to the maximum of
$v_\lambda$, and hence the selected ``velocity'' is $v_{\rm
min}(p)=v_{\lambda^*}$.  Thus, $\lambda^*$ is a solution of the
equation
\begin{equation}
\label{lambda*}
\ln\left[2(1-p)+2p\,e^{-\lambda^*}\right]
+{p\,\lambda^*\,e^{-\lambda^*}\over 1-p+p\,e^{-\lambda^*}}=0,
\end{equation}
and the selected velocity is
\begin{equation}
\label{v}
v_{\rm min}(p)={p\,e^{-\lambda^*}\over 1-p+p\,e^{-\lambda^*}}.
\end{equation}

Although it seems impossible to provide an explicit expression for 
$v_{\rm min}(p)$, one can easily probe its behavior in the limiting cases 
$p\downarrow {1\over 2}$ and $p\uparrow 1$. In these respective limits,
one gets 
\begin{equation}
\label{vp1/2}
v_{\rm min}(p)\simeq{2p-1\over \ln{1\over 2p-1}}
\end{equation}
and
\begin{equation}
\label{vp1}
1-v_{\rm min}(p)\simeq{\ln 2\over \ln{1\over 1-p}}.
\end{equation}

One can also derive the leading correction to the dominant linear
behavior of $\langle L_n^{\rm min}\rangle$,
\begin{equation}
\label{Lnlog}
\langle L_n^{\rm min}\rangle 
\simeq v_{\rm min}(p)\,n+{3\over 2\lambda^*}\,\ln n.
\end{equation}
The logarithmic correction becomes especially important when 
$p\downarrow {1\over 2}$, as can be seen from the asymptotics
\begin{equation}
\label{Ln1/2}
\langle L_n^{\rm min}\rangle \simeq 
{\epsilon\, n\over \ln(1/\epsilon)}
+{3\,\ln n\over 2\,\ln(1/\epsilon)}, \quad \epsilon=2p-1.
\end{equation}

The logarithmic correction to the front position was first derived by
Bramson in the context of a reaction-diffusion equation
\cite{bram,front1}, and was subsequently re-derived and generalized by
a number of authors\cite{B+D,van1,van2}.  Our derivation of
Eq.~(\ref{Lnlog}) follows an approach of Ref.\cite{B+D}, and we only
outline the main steps.  First, we need to examine the exact equation
(\ref{Pn}) rather than the $n\to\infty$ limit, Eq.~(\ref{Pwave}),
which was sufficient to determine the velocity.  However, the analysis
is still performed in the region far behind the front, so we can use
the linearized version of Eq.~(\ref{Pn}). Writing $P_n(x)=1-Q_n(x)$,
plugging it into Eq.~(\ref{Pn}), and ignoring quadratic terms gives
\begin{equation}
\label{Qn}
Q_{n+1}(x)=2(1-p)Q_n(x)+2pQ_n(x-1).
\end{equation}
Now we seek a solution which has the form
\begin{equation}
\label{Qnscal}
Q_n(x)=n^\alpha G\left(y\,n^{-\alpha}\right)\,e^{\lambda^* y},
\end{equation}
with $y=x-\langle L_n^{\rm min}\rangle$ as previously and yet
undetermined exponent $\alpha$ and the scaling function $G(z)$. The
correction to the leading term in the front position should be incorporated as, 
\begin{equation}
\label{front}
\langle L_n^{\rm min}\rangle \simeq v_{\rm min}(p)\,n+c(n)
\end{equation}
where the function $c(n)$ is yet undetermined.  Plugging
Eqs.~(\ref{Qnscal})--(\ref{front}) into Eq.~(\ref{Qn}) we find that
different leading orders are compatible provided that $\alpha=1/2$ and
$c(n)=\beta \ln n$ with some constant $\beta$.  Additionally, the
scaling function $G(z)$ satisfies a parabolic cylinder equation
\begin{equation}
{d^2G\over dz^2} 
+ z\,{dG\over dz} +(2\beta \lambda^*-1)G(z)=0.
\label{para}
\end{equation}
Equation (\ref{para}) should be solved subject to the boundary
conditions $G(z)\to 0$ as $z\to \infty$ as $Q_n(x)$ must vanish for
$x\to\infty$ and $G(z)\sim z$ for $z\to 0$ to ensure that $Q_n(x)$ is
independent on $n$ in the limit $n\to\infty$. The boundary condition
in the large $z$ limit selects one of the two possible solutions:
$G(z)=A\exp(-z^2/4)\,D_{2(\beta \lambda^*-1)}(z)$, where $D_\nu$ is
the parabolic cylinder function with index $\nu$.  The second boundary
condition $G(z)\sim z$ fixes the index of the parabolic cylinder
function, $2(\beta \lambda^*-1)=1$.  Hence, $\beta=3/2\lambda^*$ thus
completing the derivation of Eq.~(\ref{Lnlog}).  Note that the sign of
the logarithmic correction term in Eq.~(\ref{Lnlog}) is positive. This
is different from the most earlier studied problems\cite{B+D,van1,van2}
although the positive sign has been occasionally seen, see Ref.\cite{km}.

In the the critical case of $p=1/2$, the convergence of the
distribution $P_n(x)$ towards the asymptotic value is only algebraic,
and therefore the front propagates extremely slowly. The simplest way
to determine the rate of propagation is again to look at $P_n(x)$ far
behind the front. Writing $P_n(x)=1-Q_n(x)$ and plugging it into
Eq.~(\ref{Pn}) yields
\begin{eqnarray}
\label{Q}
Q_{n+1}(x)&=&Q_n(x)+Q_n(x-1)\nonumber\\
&-&\left[{Q_n(x)+Q_n(x-1)\over 2}\right]^2.
\end{eqnarray}
$Q_n(x)=0$ for $x\leq 0$, so the first non-trivial $Q_n$'s are
$Q_n(1)\equiv q(n)$. {}From Eq.~(\ref{Q}), $q(n+1)=q(n)-{q^2(n)\over
4}$.  In the large $n$ limit, we can employ the continuum
approximation to get ${d q\over dn}=-{q^2\over 4}$ and thence
$Q_n(1)\simeq 4n^{-1}$. Similarly, we find the asymptotics
$Q_n(2)\simeq 4n^{-1/2}$. Generally, Eq.~(\ref{Q}) implies
$Q_n(x)\simeq 2\sqrt{Q_n(x-1)}$, from which
\begin{equation}
\label{Qasympt}
Q_n(x)\simeq 4\,n^{-2^{-(x-1)}}
=4\,\exp\left(-2\,e^{\ln\ln n-x\ln 2}\right).
\end{equation}
This demonstrates that at the critical point, $Q_n(x)$ also has a
traveling wave form, $Q_n(x)=g(x-x_n)$, with the front at
$x_n\simeq (\ln 2)^{-1}\ln\ln n$. 

Equation (\ref{Qasympt}) formally applies for $x\ll x_n$. To
investigate $Q_n(x)$ in the entire range of $x$, we again use the fact
that the distribution $Q_n(x)$ should approach a traveling wave form,
$Q_n(x)\to Q(y)$ with $y=n^{-2^{-(x-1)}}$, as $n\to\infty$.  By
inserting this into (\ref{Q}) and taking $n\to \infty$ limit, we
arrive at
\begin{equation}
\label{QQ}
Q(y)=2\sqrt{Q(y^2)}-Q(y^2).
\end{equation}
{}From this equation, one finds
\begin{equation}
\label{Qsmall}
Q(y)=4y-4y^2-2y^3+...\quad {\rm for}\quad y\downarrow 0.
\end{equation}
The first term in this expansion indeed agrees with Eq.~(\ref{Qasympt}). 
Similarly, one gets
\begin{equation}
\label{Qlarge}
1-Q(y)\sim \exp\left[-{{\rm const}\over 1-y}\right] \quad
{\rm for}\quad y\uparrow 1.
\end{equation}

To provide a more rigorous derivation of the growth law for $\langle
L_n^{\rm min}\rangle$, note that $\langle L_n^{\rm min}\rangle
=\sum_{x\geq 1}P_n(x)=n-\sum_{x\geq 1}Q_n(x)$. Inserting $Q_n(x)=Q(y)$
and replacing the sum by an integral (valid for large $n$), we get,
\begin{equation}
\langle L_n^{\rm min}\rangle 
\approx n-{1\over {\ln 2}}\int_{2^{n-1}\ln n}^{\ln n} Q(e^{-z}){dz\over z}.
\label{L_n}
\end{equation}
In the limit of large $n$, the most important contribution in the
integral comes from its lower limit.  Using $Q(y)\to 1$ as $y\to 1$ (see Eq.~(\ref{Qlarge})), we arrive at
the desired result,
\begin{equation}
\langle L_n^{\rm min}\rangle \simeq (\ln 2)^{-1}\,\ln\ln n.
\label{crit}
\end{equation}

\section{Maximal Path(s)}

We now turn to the statistics of $L_n^{\rm max}$, the length of the
maximal path for the bimodal distribution.  It is now convenient to
define $R_n(x)={\rm Prob}(L_n^{\rm max}\leq x)$. By proceeding as in
the minimal length problem, it is easy to show that $R_n(x)$ evolves
according to the same equation (\ref{Pn}) with $P$ replaced by $R$, i.e.,
\begin{equation}
R_{n+1}(x)=\left[(1-p)R_n(x)+pR_n(x-1)\right]^2.
\label{R_n}
\end{equation} 
The only difference from the minimal case is in the initial
condition. Instead of Eq.~(\ref{P0}), we now have
\begin{equation}
\label{R0}
R_0(x)=\cases{1      &$x\geq 0$,\cr
              0      &$x<0$.\cr}
\end{equation}   
This difference in the initial condition, however, leads to a
different behavior of the maximal front as demonstrated below.

Not surprisingly, equation (\ref{R_n}) admits a traveling wave
solution, $R_n(x)=R(y=x-v_{\rm max}n)$ for all $p$ (see
Fig. (2)). However, in contrast to the $[1-0]$ waveform of the minimal
case, the function $R(y)$ now looks like a $[0-1]$ waveform with
$R(y)\to 0$ as a double exponential behind the front
($y\to -\infty$) and $R(y)\to 1$ exponentially ahead the front
($y\to \infty$).  We then analyze the wavefront near the
`forward' tail, as opposed to the `backward' tail of the minimal
problem.  Substituting $1-R(y)\sim e^{-\mu y}$, we find that there
exists a family of traveling wave solutions with velocity $v_{\mu}$
parametrized by $\mu$,
\begin{equation}
\label{vmu}
v_{\mu}={\ln\left[2(1-p)+2p\,e^{\mu}\right]\over \mu}.
\end{equation}   
This velocity $v_{\mu}$ has a unique minimum at $\mu=\mu^*$. By the
front selection mechanism, this minimum value $v_{\mu^*}$ is selected
as the velocity $v_{\rm max}(p)$ of the front.  Thus, we have
\begin{equation}
\label{mu*}
\ln\left[2(1-p)+2p\,e^{\mu^*}\right]
-{p\,\mu^*\,e^{\mu^*}\over 1-p+p\,e^{\mu^*}}=0,
\end{equation}
and the selected maximal velocity is
\begin{equation}
\label{vmax}
v_{\rm max}(p)={p\,e^{\mu^*}\over 1-p+p\,e^{\mu^*}}.
\end{equation}                    
Using these results together with Eqs.~(\ref{lambda*}) and (\ref{v}),
one immediately finds 
\begin{equation}
v_{\rm min}(p)+ v_{\rm max}(1-p)=1.  
\label{duality}
\end{equation}
This is the duality relation.  Unlike the behavior of the minimal
length, the maximal length does not undergo an unbinding transition
in the sense that it increases linearly with $n$ for large $n$ for
all $0\leq p\leq 1$.  However, there is still a transition at $p=1/2$:
The velocity $v_{\rm max}(p)$ increases
from $0$ to $1$ as $p$ increases from $0$ to $1/2$ and then sticks to
$1$ for $p\geq 1/2$ (see Fig. (3)).

For the maximal front also, one can easily compute the sub-leading
logarithmic correction to the front position. Proceeding as in the
minimal case, we find
\begin{equation}
\langle L_n^{\rm max}\rangle \simeq v_{\rm max}(p)\,n 
- {3\over 2\mu^*}\,\ln n.
\label{sublead}
\end{equation}
Note the negative sign in the correction term, as opposed to the
positive sign for the minimal case in Eq.~(\ref{Lnlog}). This is also
consistent with the duality relation in Eq.~(\ref{duality}). Thus,
while the coefficient $3/2$ of the logarithmic correction term seems
to be universal not just in the present problem but in many other
cases\cite{B+D,van1,van2}, the sign of the correction term is not and
can either be positive\cite{km} or negative\cite{B+D,van1,van2}.

The duality relation (\ref{duality}) can also be derived by a general
argument which does not depend on the structure of the underlying tree
and therefore is valid for extremal directed paths on arbitrary
graphs, in particular, on arbitrary lattices. The argument follows
from the observation that for bimodal distribution, if one replaces
the $0$'s by $1$'s and the $1$'s by $0$'s on the minimal path (and
thereby replacing $p$ by $1-p$), then the minimal path becomes the
maximal path. Let $n_0$ and $n_1$ denote respectively the number of
$0$'s and $1$'s on the minimal path. Evidently, $n_0+n_1=n$.  Then
clearly, $L_n^{\rm min}(p)=n_1$ and by duality, $L_n^{\rm
max}(1-p)=n_0$. Adding the two quantities, we get $L_n^{\rm
min}(p)+L_n^{\rm max}(1-p)=n$. Dividing by $n$, we immediately get the
duality relation Eq.~(\ref {duality}).  Below, we shall derive a more
generalized version of this duality relation which goes beyond the
bimodal distribution.

\section{Generalization to Trees with Random Branching}

The generalization of the above analysis to the case of the rooted
Cayley tree with a random number of branches is straightforward.  Let
$b_m$ is the probability that the number of branches is equal to $m$
and $b$ is the average number of branches, $b=\sum_{m\geq 1} mb_m$.
Equation (\ref{Pn}) is replaced by
\begin{equation}
\label{Pnb}
P_{n+1}(x)=\sum_{m=1}^\infty b_m\left[(1-p)P_n(x)+pP_n(x-1)\right]^m,
\end{equation}
and the subsequent analysis repeats the steps detailed in the case of
the binary tree. In particular, $P_n(x)$ approaches the stationary 
distribution for $p<p_c$, while above $p_c$ the minimal length grows
according to Eq.~(\ref{Lnlog}) with $\lambda^*$ a solution to equation
\begin{equation}
\label{lb}
\ln\left[b(1-p)+bp\,e^{-\lambda^*}\right]+
{p\,\lambda^*\,e^{-\lambda^*}\over 1-p+p\,e^{-\lambda^*}}=1,
\end{equation}
and $v_{\rm min}(p)$ given by the same relation (\ref{v}).  At the
critical point $p_c=1-b^{-1}$, the minimal length is given by the same
expression (\ref{crit}) for any tree. Also, the velocity of the
maximal path can be determined from the general duality relation in
Eq.~(\ref{duality}).

\section{Generalization to Arbitrary Distributions}

We now consider the statistics of extreme path(s) on a binary tree for
arbitrary distribution $\rho(l)$ of the edge lengths. Let us first
consider the minimal path(s).  The velocity of the minimal path for an
arbitrary distribution can be extracted from earlier results by Derrida and
Spohn\cite{D+S} which we re-derive below for the sake of completeness.
We consider again $P_n(x)={\rm Prob}(L_n^{\rm min}\geq x)$. Proceeding
as in the case of the bimodal distribution, one finds that the appropriate
generalization of Eq.~(\ref{Pn}) is given by
\begin{equation}
P_{n+1}(x)=\left[\int_0^{\infty}dl\, \rho(l) P_n(x-l)\right]^2.
\label{P_ng}
\end{equation}
with the same initial condition, Eq.~(\ref{P0}), as earlier. The
distribution of the minimal length is then given by $P_n^{\rm
min}(x)=-dP_n(x)/dx$ and the average minimal length is $\langle
L_n^{\rm min}\rangle=\int_0^{\infty}P_n(x)dx$.

Defining $P_n(x)=G_n^2(x)$, we recast Eq.~(\ref{P_ng}) into
\begin{equation}
G_{n+1}(x)=\int_0^{\infty}dl\, \rho(l) G_n^2(x-l).
\label{G_ng}
\end{equation}  
For any given $n$, $G_n(x)\to 1$ as $x\to -\infty$ and $G_n(x)\to 0$
as $x\to \infty$. Substituting $G_n(x)=1-g_n(x)$ in Eq. (\ref{G_ng})
and using normalization of $\rho(l)$ one finds,
\begin{equation}
g_{n+1}(x)=\int_{-\infty}^x dy\, \rho(x-y)\left[ 2g_n(y)-g_n^2(y)\right],
\label{g_ng}
\end{equation}
where $g_n(x)\to 0$ as $x\to -\infty$ and $g_n(x)\to 1$ as $x\to
+\infty$. We analyze the above equation in the `backward' tail region,
i.e., when $x\to -\infty$. In this limit, one can neglect the
nonlinear term in $g_n$ inside the integral in Eq.~(\ref{g_ng}). The
resulting linear equation admits a traveling wave solution of the
form, $g_n(x)=e^{\lambda(x-v_{\lambda}n)}$, with
\begin{equation}
v_{\lambda}=-{1\over {\lambda}}\,
\ln\left[ 2\int_0^{\infty}dz\, e^{-\lambda z}\rho(z)\right].
\label{vg}
\end{equation}
For generic length distributions $\rho(l)$, this function has a
maximum at $\lambda=\lambda^*$ and by the general velocity selection
principle, this maximum velocity is selected, i.e., $v_{\rm
min}=v_{\lambda^*}$.

The question we are interested in is for which class of distributions
$\rho(l)$, an unbinding transition can occur for the minimal
path. Such a transition will occur if the velocity $v_{\lambda}$ in
Eq.~(\ref{vg}) ceases to have a unique maximum. By analyzing
Eq.~(\ref{vg}) one sees that this can happen only if $\rho(l)$ has a
nonzero delta-function weight at $l=0$, i.e., when $\rho(l)=a\delta(l)
+ (1-a)f(l)$ with $0<a<1$ and $f(l)$ is some positive function with
$\int_0^{\infty}dl\, f(l)=1$. The unbinding transition occurs as the
parameter $a$ is varied. Note that the positivity condition of
velocity in Eq.~(\ref{vg}) demands that $2a<1$. Thus the critical
point always occurs at $a=1/2$.  The bimodal distribution considered
in the previous sections is a special case of this class of
distributions with $a=1-p$ and $f(l)=\delta(l-1)$.
 
For $a<1/2$, the average minimal length increases linearly with $n$
for large $n$ with the velocity $v_{\lambda^*}$ obtained from
Eq. (\ref{vg}). For $a>1/2$, the function $P_n(x)$ reaches a
stationary distribution for large $n$ and hence the $\langle L_n^{\rm
min}\rangle$ saturates to a nonzero constant as $n\to \infty$. What
happens at the critical point $a=1/2$ for generic distributions
$f(l)$'s? For $a=1/2$, the Eq. (\ref{G_ng}) reduces to,
\begin{equation}
G_{n+1}(x)={1\over 2}G_n^2(x)+ 
{1\over 2}\int_0^{\infty}dl\, f(l) G_n^2(x-l).
\label{G_ngc}
\end{equation}    
Detailed analysis of the above equation reveals a rather remarkable
universal property at the critical point.  It turns out that for
generic $f(l)$, there are only two possible behaviors of $\langle
L_n^{\rm min}\rangle =\int_0^{\infty} dx\,G_n^2(x)$, depending on whether
the function $f(l)$ has a gap or not near $l=0$. If $f(l)$ does not
have a gap at $l=0$, it turns out that as $n\to \infty$, $G_n(x)$ in
Eq. (\ref{G_ngc}) tends to a stationary distribution and hence
$\langle L_n^{\rm min}\rangle \to {\rm const}$. In the opposite case,
when $f(l)$ has a gap $\Delta$ near $l=0$, it turns out that as $n\to
\infty$,
\begin{equation}
\langle L_n^{\rm min}\rangle\simeq {\Delta\over {\ln 2}}\ln\ln n .  
\label{lnlng}
\end{equation}
The bimodal distribution corresponds to the special case $\Delta=1$.

These general results are best demonstrated by specific examples which
can be worked out explicitly. Let us first consider an example which
is gapless, such as the exponential distribution, $f(l)=e^{-l}$.
Remarkably, Eq.~(\ref{G_ngc}) can be solved exactly in this case. It
is easy to verify that Eq.~(\ref{G_ngc}) admits an exponential
solution, $G_n(x)=A_n e^{-x}$, where $A_{n+1}=(1+A_n^2)/2$ with
$A_0=0$. As $n\to \infty$, $A_n$ approaches to $1$ (more precisely,
$A_n\simeq 1-2/n$). Therefore, $\langle L_n^{\rm min}\rangle=\int
dx\,G_n^2(x)=A_n^2/2$ approaches to 1/2 as $n\to \infty$.  Next we
consider a distribution with gap such as $f(l)={1\over
4}\delta(l-1)+{1\over 4}\delta(l-2)$. This case also can be worked out
explicitly following the same steps as used for the bimodal case. It
turns out that essentially all the steps are identical to the bimodal
case, except that $n$ in Eq. (\ref{crit}) gets replaced by $n/2$.  But
this does not change the leading asymptotic behavior for large $n$
which is still given by Eq. (\ref{crit}).

We now turn to the maximal path(s) for an arbitrary distribution
$\rho(l)$. As in the case of the bimodal distribution, $R_n(x)={\rm
Prob}(L_n^{\rm max}\leq x)$ satisfies the same equation (\ref{P_ng})
as $P_n(x)$, i.e., 
\begin{equation} 
R_{n+1}(x)=\left[\int_0^{\infty}dl\,
\rho(l) R_n(x-l)\right]^2, 
\label{R_ng} 
\end{equation} 
the only difference is that the initial condition, Eq.~(\ref{P0}), 
should be replaced by Eq.~(\ref{R0}).  Substituting 
$R_n(x)=[1-s_n(x)]^2$, we get
\begin{equation} 
s_{n+1}(x)=\int_{-\infty}^x dy\, \rho(x-y)\left[
2s_n(y)-s_n^2(y)\right], 
\label{s_ng} 
\end{equation} 
with the boundary conditions $s_n(x)\to 1$ as $x\to -\infty$ and
$s_n(x)\to 0$ as $x\to \infty$. We now have to analyze the `forward'
tail of the front, i.e., the behavior in the $x\to \infty$ limit. We
first consider bounded distributions $\rho(l)$'s with an upper cut-off
$\Lambda$. In that case, the lower limit of the integration in
Eq.~(\ref{s_ng}) becomes $x-\Lambda$. In the tail region where $x\gg
\Lambda$, one can again neglect the nonlinear term in $s_n$ inside the
integral in Eq.~(\ref{s_ng}) and the resulting linear equation admits
a traveling wave solution of the form $s_n(x)=e^{-\mu(x-v_\mu n)}$,
where
\begin{equation} 
v_{\mu}={1\over {\mu}}\log\left[ 2\int_0^{\Lambda}dz\,
e^{\mu z}\rho(z)\right].  
\label{vs} 
\end{equation} 
For generic distribution, $v_{\mu}$ has a unique minimum at
$\mu=\mu^*$ and via the front selection principle, this minimum
velocity is chosen and one gets, $v_{\rm max}=v_{\mu^*}$.  Comparing
Eqs.~(\ref{vg}) and (\ref{vs}) it becomes evident that for the bounded
{\em symmetric} distributions, i.e., when $\rho(l)=\rho(\Lambda-l)$,
one gets the general relation,
\begin{equation} 
v_{\rm min}+v_{\rm max}=\Lambda.
\label{gendual} 
\end{equation}

For unbounded distributions, however, one has to be careful since it
is not obvious that one can neglect the nonlinear term in $s_n$ inside
the integral in Eq. (\ref{s_ng}).  Equation (\ref{vs}) for the
maximal velocity still remains valid as long as $\rho(l)$ decays with
$l$ exponentially or faster. To see this explicitly in an example, let
us consider the exponential distribution, $\rho(l)=e^{-l}$. In this
case, one can transform the the integral equation (\ref{s_ng}) into
the following difference-differential equation,
\begin{equation}
{ds_{n+1}(x)\over {dx}}=2s_n(x)-s_n^2(x)-s_{n+1}(x).
\label{exps}
\end{equation}  
In this differential form, it is clear that for large $x$, one can 
neglect the nonlinear term on the right-hand side of
Eq.~(\ref{exps}).  By inserting $s_n(x)\sim e^{-\mu(x-v_{\mu}n)}$ 
into resulting linear equation, one gets
\begin{equation}
v_{\mu}={1\over {\mu}}\,\ln\left({2\over {1-\mu}}\right).
\label{expv}
\end{equation}
Note that the same formula is obtained if one substitutes
$\rho(l)=e^{-l}$ directly into the general expression
(\ref{vs}). Minimizing with respect to $\mu$, one gets $v_{\rm max}$.
The corresponding minimal velocity $v_{\rm min}$ for the exponential
distribution can be obtained from the general formula in
Eq.~(\ref{vg}) with $\rho(z)=e^{-z}$ and then maximizing with respect
to $\lambda$. So, for the exponential distribution we finally get,
\begin{equation}
v_{\rm min}=0.231961\ldots \quad {\rm and}\quad v_{\rm max}=2.678346\ldots
\label{expv2}
\end{equation}
Thus for unbounded distributions, there does not seem to exist any 
simple relation between minimal and maximal velocities.

\section{Conclusion}

In this work, we have shown that the length of the minimal path on a
Cayley tree exhibits an unbinding phase transition from a localized
phase to a moving phase.  This phase transition is driven by the
parameter $p$ of the bimodal distribution.  In the localized phase
($p<p_c$), the minimal length distribution approaches a stationary
depth-independent form $P^{\rm min}(x)$. This distribution vanishes
extremely sharply, as the double exponential, in the large $x$ limit.
In the moving phase ($p>p_c$), the minimal length distribution
$P_n^{\rm min}(x)$ approaches a traveling wave form with the front
velocity relaxing algebraically, $v_{\rm min}=v(p)+3/(2\lambda^*
n)+{\cal O}\left(n^{-3/2}\right)$, in the large $n$ limit.
Specifically, $P_n^{\rm min}(x)$ approaches the {\em solitary}
traveling wave ${\cal P}(y)-{\cal P}(y+1)$, with $y=x-\langle
L_n^{\rm min}\rangle$, the front position $\langle L_n^{\rm min}\rangle$ given by
Eq.~(\ref{Lnlog}), and ${\cal P}(y)$ being a solution of
Eq.~(\ref{Pwave}).  In the critical case, $p=p_c$, the minimal length
distribution approaches the solitary traveling wave, although the
front propagates extremely slowly, as an iterated logarithm.  Given
that the distribution of the minimal length is the soliton of a finite
width, the variation of the minimal length is finite. In other words,
$L_n^{\rm min}$ should not vary much from sample to sample. 

We have also studied the complementary problem of the statistics of 
the maximal path. We have found that the maximal length always grows
linearly with the height, with the velocity $v_{\rm max}$ again determined via
the front selection mechanism.  For the bimodal length distribution,
the minimal and maximal velocities are connected via the duality relation, 
$v_{\rm min}(p)+v_{\rm max}(1-p)=1$, which admits extensions to
finite-dimensional situations and to arbitrary {\em bounded symmetric}
length distributions. 

We have found that the tail of the minimal length distribution is a
double exponential, in contrast with a simple exponential that occurs
in most traveling wave problems\cite{front2}.  This is not very
surprising since similar tails were found in statistics of
extremes\cite{extreme}.  Note, however, that our exact results
exemplify extreme value statistics for {\em correlated} random
variables (see \cite{C+D} for a recent review), while classical
results\cite{extreme} correspond to the case of {\em independent}
identically distributed random variables.

\medskip
\noindent
PLK acknowledges support from NSF, ARO, and CNRS.

\end{multicols}

\begin{figure}
\begin{center}
\leavevmode
\psfig{figure=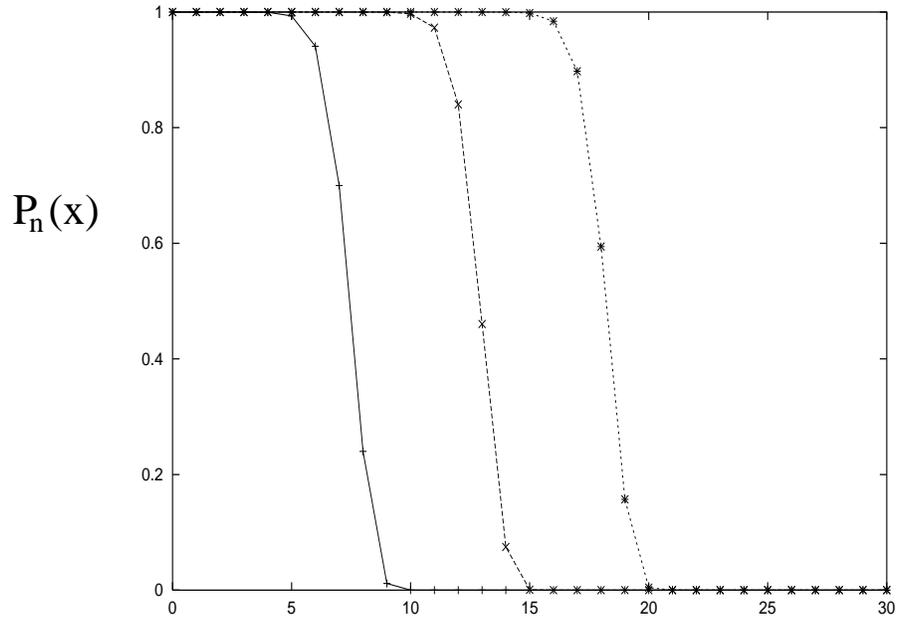,width=12cm,angle=0}
\caption{The propagating front for the cumulative distribution $P_n(x)$ 
of the minimal length for the bimodal distribution with $p=0.8$.}
\end{center}
\end{figure}       

\begin{figure}
\begin{center}
\leavevmode
\psfig{figure=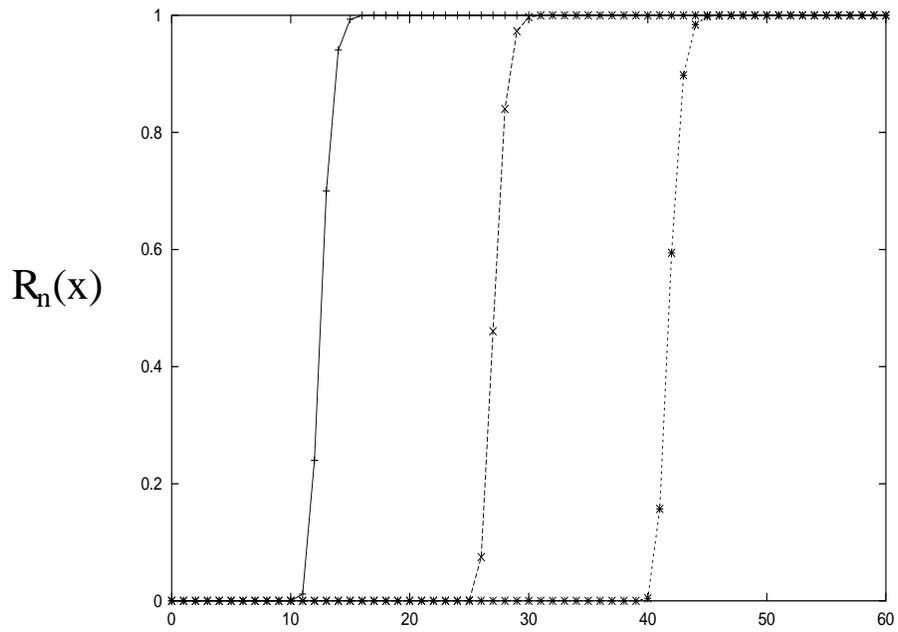,width=12cm,angle=0}
\caption{The propagating front for the cumulative distribution $R_n(x)$ 
of the maximal path for the bimodal distribution with $p=0.2$.}
\end{center}
\end{figure}

\begin{figure}
\begin{center}
\leavevmode
\psfig{figure=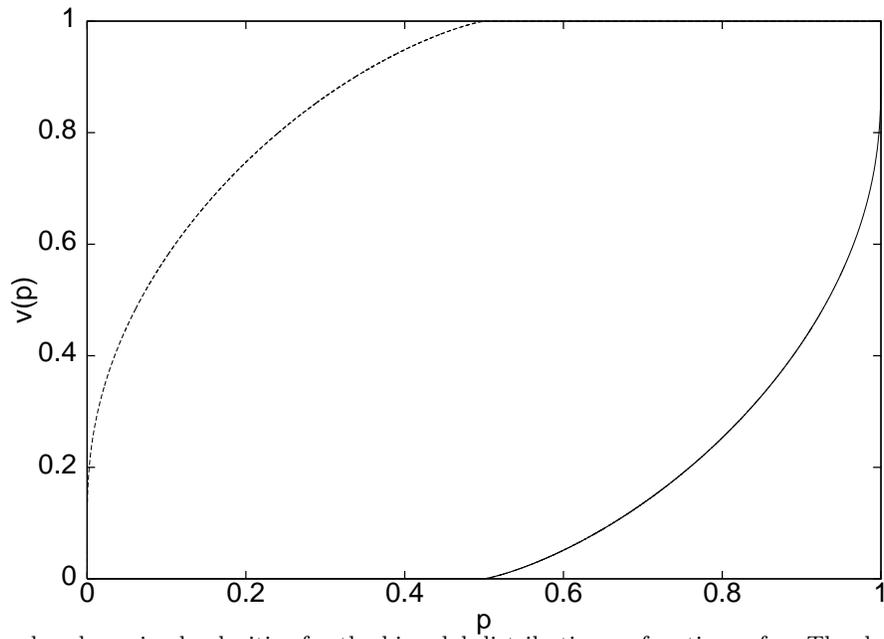,width=12cm,angle=-90}
\caption{The minimal and maximal velocities for the bimodal distribution 
as functions of $p$. The dotted line
shows $v_{\rm max}(p)$ and the solid line represents $v_{\rm min}(p)$.}
\end{center}
\end{figure}

\end{document}